\documentclass[aps,prd,superscriptaddress,tightenlines,preprintnumbers,nofootinbib,11pt]{revtex4-1}
\pdfoutput=1

\usepackage[english]{babel}
\usepackage{amsmath,amssymb,amsbsy,amstext,amsthm,booktabs,simplewick,exscale,relsize,slashed,graphicx,amsfonts,upgreek}
\usepackage{tabu,multirow,color,dcolumn,bm,enumerate}
\usepackage{hyperref}

\newcommand{\tev}{\text{TeV}}
\newcommand{\gev}{\text{GeV}}
\newcommand{\fbinv}{\text{fb}^{-1}}
\newcommand{\abinv}{\text{ab}^{-1}}
\newcommand{\eff}{\text{eff}}
\usepackage{graphicx,bm}
\usepackage[svgnames]{xcolor}

\newcommand{\beq}{\begin{equation}}
\newcommand{\bea}{\begin{eqnarray}}
\newcommand{\eeq}{\end{equation}}
\newcommand{\eea}{\end{eqnarray}}
\newcommand{\bal}{\begin{align}}
\newcommand{\eal}{\end{align}}

\usepackage{tikz}
\usetikzlibrary{decorations.pathmorphing,decorations.markings}
\tikzset{
photon/.style={decorate, decoration={snake,amplitude=4pt, segment length=7pt}, draw=black},
particle/.style={draw=black, postaction={decorate}, decoration={markings,mark=at position .5 with {\arrow[draw=black]{>}}}},
antiparticle/.style={draw=black, postaction={decorate}, decoration={markings,mark=at position .5 with {\arrow[draw=black]{<}}}},
gluon/.style={decorate, draw=black, decoration={coil,amplitude=3pt, segment length=4pt}},
higgs/.style={draw=black,dashed,thick },
arrow/.style={draw=black, very thick, postaction={decorate}, decoration={markings,mark=at position 1 with {\arrow[draw=black]{>}}}}
}
\begin{document}

\title{Boosted Higgses from chromomagnetic $b$'s: $b\bar{b}h$ at high luminosity}
\author{Joseph Bramante\footnote{\tt jbraman2@nd.edu}}
\affiliation{Department of Physics, University of Notre Dame, 225 Nieuwland Hall, Notre Dame, IN, USA, 46556}
\author{Antonio Delgado\footnote{\tt antonio.delgado@nd.edu}} 
\affiliation{Department of Physics, University of Notre Dame, 225 Nieuwland Hall, Notre Dame, IN, USA, 46556}
\affiliation{PH-TH Department, CERN, CH-1211, Geneva 23, Switzerland}
\author{Landon Lehman\footnote{\tt llehman@nd.edu}}
\affiliation{Department of Physics, University of Notre Dame, 225 Nieuwland Hall, Notre Dame, IN, USA, 46556}
\author{Adam Martin\footnote{\tt amarti41@nd.edu}}
\affiliation{Department of Physics, University of Notre Dame, 225 Nieuwland Hall, Notre Dame, IN, USA, 46556}

\preprint{CERN-PH-TH/2014-179} 

\begin{abstract}
\vspace*{0.5cm}

	This paper examines detection prospects and constraints on the
	chromomagnetic dipole operator for the bottom quark. This operator has a
	flavor, chirality and Lorentz structure that is distinct from other
	dimension six operators considered in Higgs coupling studies. Its
	non-standard Lorentz structure bolsters boosted $b \bar{b} h$ events,
	providing a rate independent signal of new physics. To
	date, we find this operator is unconstrained by $p p \rightarrow
	h + {\rm jets}$ and $pp \rightarrow \bar b b $ searches: for order-one
	couplings the permitted cutoff $\Lambda$ for this operator can
	be as low as $\Lambda \sim 1~{\rm TeV}$. We show how to improve this
	bound with collider cuts that allow a $b$-tagged 
	Higgs plus dijet search in the Higgs to diphoton decay channel to
	exclude cutoffs as high as $\sim 6~{\rm TeV}$ at $2 \sigma$ with 3
	$\abinv$ of luminosity at the 14 TeV LHC. Cuts on the $p_T$ of the Higgs
	are key to this search, because the chromomagnetic dipole yields a
	non-standard fraction of boosted Higgses.
\end{abstract}

\maketitle

\tableofcontents

\section{Introduction}
\label{sec:intro}

The bottom quark is the heaviest fundamental fermion lighter than the Higgs, the predominant
	decay product of the Higgs, and the most easily tagged quark at LHC
	energies. For these reasons, new physics contributions from effective operators with bottom quarks will be sought out as the LHC ramps to higher energies. The search is already afoot for non-standard
Higgs interactions encapsulated in higher-dimensional Higgs operators
\cite{Burges:1983zg,Leung:1984ni,Buchmuller:1985jz,DeRujula:1991se,Hagiwara:1993qt,GonzalezGarcia:1999fq,Eboli:1999pt,Giudice:2007fh,Grzadkowski:2010es,Espinosa:2012im,Azatov:2012qz,Low:2012rj,Ellis:2012hz,
Alloul:2013naa,Brivio:2013pma,Englert:2014uua}. Effective operator searches that
are sensitive to a subset of possible operators are particularly important,
because each independent high energy cross-section
measurement of Higgs couplings provides information that complements constraints
from flavor, precision electroweak, and also non-Higgs searches at the Tevatron
and LHC. For certain higher dimensional operators, high energy Higgs studies
will place the strongest bounds, or more optimistically, the best opportunity
for discovery. 

Motivated by these considerations, this article examines non-standard interactions between the Higgs and bottom
quarks. Bottom quarks play the most important role in Higgs
physics since $\Gamma(h \to \bar b b)$ is the largest partial width of the Higgs boson. Despite
this, alternative Higgs-bottom dynamics are relatively unexplored. Existing work
\cite{Hayreter:2013kba} that does incorporate non-Standard Model (SM) $hb\bar
b$ dynamics has thus far focused on SM-like (i.e.~Yukawa) interactions with
non-SM strength, rather than more general kinematic structures.

New physics operators can affect Higgs observables in two ways: normalization
and shape. By normalization we mean that adding new physics affects the overall
rate of events, but does not change any differential kinematic distributions,
while a shape change means that the total event rate is unchanged but the kinematic
distributions of particles shift. For example, consider the SM extended by a new
physics (NP) operator
\begin{equation}
\mathcal L_1 \supset c_{SM}\, \mathcal O_{SM} + c_{NP}\, \mathcal O_{NP}. 
\end{equation}
If the Lorentz or chirality structure and field content of the two operators is the
same, the NP effects can be recast as a change in the SM coefficient: $c_{SM} \to
c_{SM} + f(c_{NP})$. In this case, all distributions will be SM-like but the
total rate will change. Most Higgs constraints to date have focused on this
possibility. However, if the two operators have different Lorentz structures, the NP and
SM events can be distinguished by using kinematic cuts to select for non-standard
distributions of final state particles.

This distinction between normalization effects and shape effects can also be
phrased in terms of signal strength $\mu^{\text{collider}}$, defined as the
ratio of Higgs events in some new physics scenario relative to the number of
events in the SM. The number of events is the product of the luminosity
$\mathcal L$, the production cross section $\sigma(pp \to h+X)$ (where X is some
other potential SM particle produced in association with the Higgs), the Higgs
branching ratio $BR$ into whatever final state we are observing, the analysis
cut acceptance $A$, and the efficiency $\epsilon$. Throughout this article we
make a distinction between the acceptance, which is the fraction of events that
pass certain cuts, and the efficiency, the probability that objects satisfying
the cuts are correctly captured by the experiment. Written out explicitly
\begin{equation}
\mu^{\text{collider}} = \frac{\mathcal L\, \sigma(pp \to h + X)\, BR\, A\,\epsilon}{\mathcal L\, \sigma(pp \to h + X)_{SM}\, BR_{SM}\, A_{SM}\,\epsilon}
\end{equation}
The luminosity and efficiency cancel in the ratio, but the acceptance does not.
The only time the acceptance cancels in the ratio is when new physics modifies
the strength of the Higgs interaction, but does not change the kinematics of
events, leaving acceptance unchanged; in this case, the signal strength reduces
to the ratio of production cross sections times branching ratios, what we will
call $\mu^{\text{parton}}$,
\begin{equation}
\mu^{\text{parton}} = \frac{\sigma(pp \to h + X)\, BR}{\sigma(pp \to h + X)_{SM}\, BR_{SM}}.
\end{equation}
This simplified scenario is attractive because it contains only theoretically
defined quantities and is independent of the experiment. However, it only
applies to a limited set of new physics scenarios. In general the acceptance
must be included and the signal strength becomes a more involved and
analysis-dependent quantity.  

The $b$ quark chromomagnetic dipole operator, defined in the next section and examined throughout this
paper, is one example of an operator which changes both the rate and kinematic
distribution of collider events which contain a Higgs boson, making it ideal for
a study of new physics arising from non-standard Higgs couplings to $b$ quarks. 
In this paper we explore existing constraints on and high luminosity discovery
prospects for the bottom quark chromomagnetic dipole operator.  While our focus
will be on the Standard Model augmented with a single new physics operator, the
methods and
constraints herein can also be applied to more complicated scenarios involving
multiple operators. Non-standard kinematic structure is not unique to the
chromomagnetic $b$-quark operator, and several studies exist in the literature
exploring how kinematic distributions can be used to pin down certain new
physics effects in associated production $pp \to W/Z + h$, Higgs plus jet
production $pp \to h + j$, and $pp \to t\bar t h$ events~\cite{Soper:2011cr,
Almeida:2011aa,Katz:2010iq,Degrande:2012gr,Corbett:2012dm,Corbett:2012ja,Ellis:2012xd,Isidori:2013cla,Isidori:2013cga,Harlander:2013oja,Banfi:2013yoa,Corbett:2013pja,Azatov:2013xha,Brivio:2013pma,Grojean:2013nya,Anders:2013oga,Plehn:2013paa,Englert:2013vua,deLima:2014dta,Belanger:2014roa,Bramante:2014gda,Schlaffer:2014osa}.

The remainder of this paper is organized as follows. In Section \ref{sec:bchrome}, we
comment on the properties and structure of the $b$ quark chromomagnetic dipole
operator $\mathcal O_{ghd}$,
pointing out the kinematic and chiral properties it has which differ from the
SM. Section \ref{sec:constraints} surveys the interactions from
$\mathcal O_{ghd}$ that lead to new physics contributions to $pp \to \bar b b h$
and $p p \to \bar b b$ at tree level, and to $pp \to h$ at one loop level, then
addresses the constraints from existing searches in each of these channels. In
Section \ref{sec:btagit} we discuss opportunities to explore BSM $b \bar b h$
parameter space with improved searches at both the 8 and 14 $\tev$ LHC runs, both through b-tagging
and selection of Higgs decays boosted by chromomagnetically dipolarized $b$ quarks. We conclude in
Section \ref{sec:conc} with a discussion of high luminosity prospects for
the bottom quark chromomagnetic dipole.

\section{The chromomagnetic dipole}
\label{sec:bchrome}

To pursue a clear example of non-standard final state kinematic morphology
for $b$ quarks and the Higgs boson, in this article we focus on the bottom
quark chromomagnetic dipole operator
\begin{equation} 
\mathcal O_{ghd} =
\frac{c_{ghd}}{\Lambda^2_0}\,(Q^{\dag}_j H)\,Y_{d,ij}\,\overline{\sigma}_{\mu\nu}t^{A}\,
d^{c\dag}_i\, G_A^{\mu\nu} + h.c.
\label{eq:ghd}
\end{equation}
Here $c_{ghd}$ is the new physics coupling, $\Lambda_0$ is the energy scale
suppressing the higher-dimensional operator, and $Y_d$ is an insertion of the
down-type quark Yukawa matrix. There are a number of works that list dimension-6
operators and constrain them with Higgs measurements (see
Refs.~\cite{Einhorn:2013tja,Ellis:2014dva,Englert:2014uua}) and much recent work
studying the coupling of the Higgs to third generation quarks
\cite{Brod:2013cka,Bevilacqua:2014qfa,Khatibi:2014bsa,Schlaffer:2014osa,Demartin:2014fia,Dawson:2014ora,Wiesemann:2014ioa}.
However, the $b$ quark chromomagnetic dipole has only been studied for its
modification of the $pp \rightarrow h+X$ cross-section (where X is another SM parton)
\cite{Hayreter:2013kba}. Some work, both pre- and post-Higgs discovery, on the
collider bounds on the top quark chromomagnetic dipole, the cousin of
Eq.~(\ref{eq:ghd}), can be found in
\cite{Degrande:2012gr,Hayreter:2013kba,Bramante:2014gda}.

By including the down-quark Yukawa matrix in Eq.~(\ref{eq:ghd}) we have rendered this operator automatically
minimally-flavor-violating, meaning the same field redefinitions that diagonalize
the down-quark mass matrix also diagonalize $\mathcal O_{ghd}$. This guarantees that
the $\mathcal O_{ghd}$ operator does not induce new flavor-violating interactions
(at least at tree-level), which are tightly constrained by flavor physics \cite{Isidori:2010kg}.
Diagonalized, $\mathcal O_{ghd}$ contains chromomagnetic dipole interactions for
the down and strange quarks, as well as for the bottom. However, as the down and
strange interactions are suppressed by their corresponding tiny Yukawa
couplings, we will forget about them here. We will also assume that $c_{ghd}$ is real to avoid constraints from CP-violation. Thus, Eq.~(\ref{eq:ghd}) approximately reduces to a chromomagnetic dipole moment for the bottom quark alone
\begin{align}
\mathcal O_{ghd} & \sim
\frac{c_{ghd}}{\Lambda^2_0}\,(Q_3^{\dag}H)\,y_b\,\overline{\sigma}_{\mu\nu}t^{A}\,
b^{c\dag}\, G_A^{\mu\nu} + h.c.
\label{eq:bchrome}
\end{align}

Our main interest in this paper is the constraints and prospects for the
operator in Eq.~(\ref{eq:bchrome}), however there are multiple ways that we can
express its strength. One possibility is to combine the $y_b$ factor,
coefficient $c_{ghd}$ and $\Lambda_0$ all into an `effective cutoff'
$\Lambda_{\eff}$, which we  can adjust. Another possibility is to fix
$\Lambda_0$ to some value, then separately dial $c_{ghd}$. Within the effective
theory, this difference in presentation is purely aesthetic. The convention we
will use throughout this paper is to fix the combination
\begin{align}
\frac{y_b}{\Lambda^2_0} \equiv \frac{1}{\Lambda^2} =  \frac{1}{(6\,\tev)^2}
\label{eq:masscouplconv}
\end{align}
then vary $c_{ghd}$ to adjust the strength of $\mathcal O_{ghd}$. We choose
this convention for two reasons: i.) for $\Lambda_0 = 1\,\tev$, the same value
we used when studying the chromomagnetic top quark operator in
Ref.~\cite{Bramante:2014gda}\footnote{In that work, as $y_t \cong 1$, we did
not distinguish between $\Lambda_0$ and $\Lambda$.}, the combination
$y_b/\Lambda^2_0 \sim 1/(6\,\tev)^2$, and ii.) the highest dijet invariant
masses currently probed by the LHC are $\sim 6\,\tev$. As we will see,
$\mathcal O_{ghd}$ contributes to $pp \to \bar b b$ production, so a natural
starting point for the scale suppressing $\mathcal O_{ghd}$ is the highest
scale probed in $pp \to b\bar b$. In addition to the ambiguities in defining
the strength of $\mathcal O_{ghd}$, there are also subtleties in its
interpretation, especially for $c_{ghd} > 1$, which we comment on in
Sec.~\ref{sec:otherops}.

\begin{figure}
\begin{tabular}{c|c}
\includegraphics[scale=1.5]{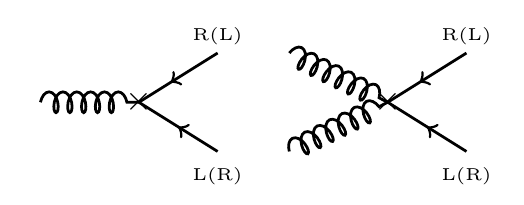}&\includegraphics[scale=1.5]{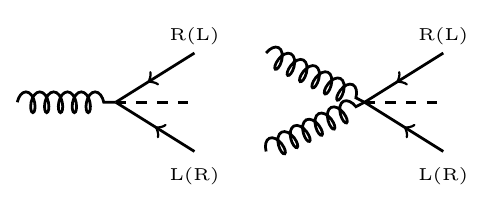} \\
(A) & (B)
\end{tabular}
\caption{(A) These
	diagrams illustrate that the interference term between SM and chromomagnetic
	dipole $b\bar{b}$ production will require a chirality flip. This flip suppresses the
	bottom quark chromomagnetic dipole contribution to bottom diquark production
	(relative to the top quark, which receives substantial corrections to $t
	\bar{t}$ production from its chromomagnetic dipole) by a factor of the bottom
	Yukawa coupling, $y_b$. (B) These diagrams show $s$-channel $b \bar{b} h$ processes from $\mathcal{O}_{ghd}$.}
\label{fig:bbhdiag}
\end{figure}

Expanding the Higgs about its vacuum expectation value, $\mathcal O_{ghd}$
contains four separate interactions: a 4-point interaction involving a Higgs,
gluon, $b_L$ and $b_R$, a 5-point interaction with a Higgs, two gluons, $b_L$
and $b_R$,  a 4-point interaction involving two gluons, $b_L$ and $b_R$,  and a
3-point interaction between a gluon, $b_L$ and $b_R$. Each of these $n$-point
interactions introduces detectable modifications to the SM. The Feynman diagrams
for these vertices are
displayed in Fig. \ref{fig:bbhdiag}. All interactions involve
opposite chirality $b$ quarks, and the single gluon interactions are all
proportional to the momentum carried by the gluon. These two features are the
source of the kinematic difference between $\mathcal O_{ghd}$ and the SM, and
the subject of the next subsections. 

\subsection{Boosted particles from the chromomagnetic dipole}
\label{sec:bboost}

\begin{figure}
\begin{tabular}{c}
\includegraphics[scale=1.5]{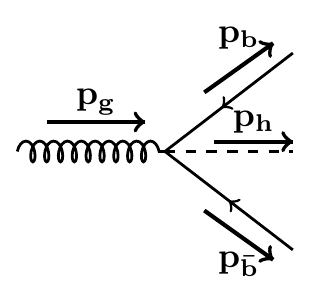}
\end{tabular}
\caption{This diagram shows $s$-channel gluon production of $b \bar{b} h$. Note that, in contrast to SM processes where the Higgs must be radiated from a fermion, here the amplitude is proportional to the gluon momentum, which scales directly with the momentum of the Higgs: ${\bf p_g = p_b + p_{\bar{b}} + p_h}$. The process $pp \to b\bar b$ proceeds through the same diagram, with the Higgs set to its vacuum expectation value; there ${\bf p_g = p_b + p_{\bar{b}}}$.}
\label{fig:boosted}
\end{figure}

The first structural feature we address is the gluon momentum dependence carried
by all single-gluon interactions in  $\mathcal O_{ghd}$. This feature will make
the biggest impact in processes where the gluon momentum can be large, and
little to no impact in processes where the gluon momentum is fixed and small.
Two processes that sit in the first category  are $pp \to b \bar b h$ and $ p p
\to b \bar b$. Both contain $s$-channel gluon diagrams with one end of the gluon
propagator terminating in the momentum-dependent $\mathcal O_{ghd}$ vertex (see
Fig.~\ref{fig:boosted}). The larger the $\sqrt{\hat s}$, or, equivalently, the
larger the boost of the final state $b, \bar b$ or $h$, the larger the effect
$\mathcal O_{ghd}$ has on the process. This trend can be contrasted with $pp \to
h$, which receives a contribution from the momentum dependent $\mathcal O_{ghd}$
interactions at one-loop level (see Fig.~\ref{fig:higgsloop}). In $pp \to h$,
the characteristic momentum scale of the incoming gluons is fixed to $O(m_h) \ll
\Lambda$, so there is no way to enhance the new physics effects by cutting
events to find special corners of kinematic phase space.

\begin{figure}[h!]
\includegraphics[width=0.45\textwidth]{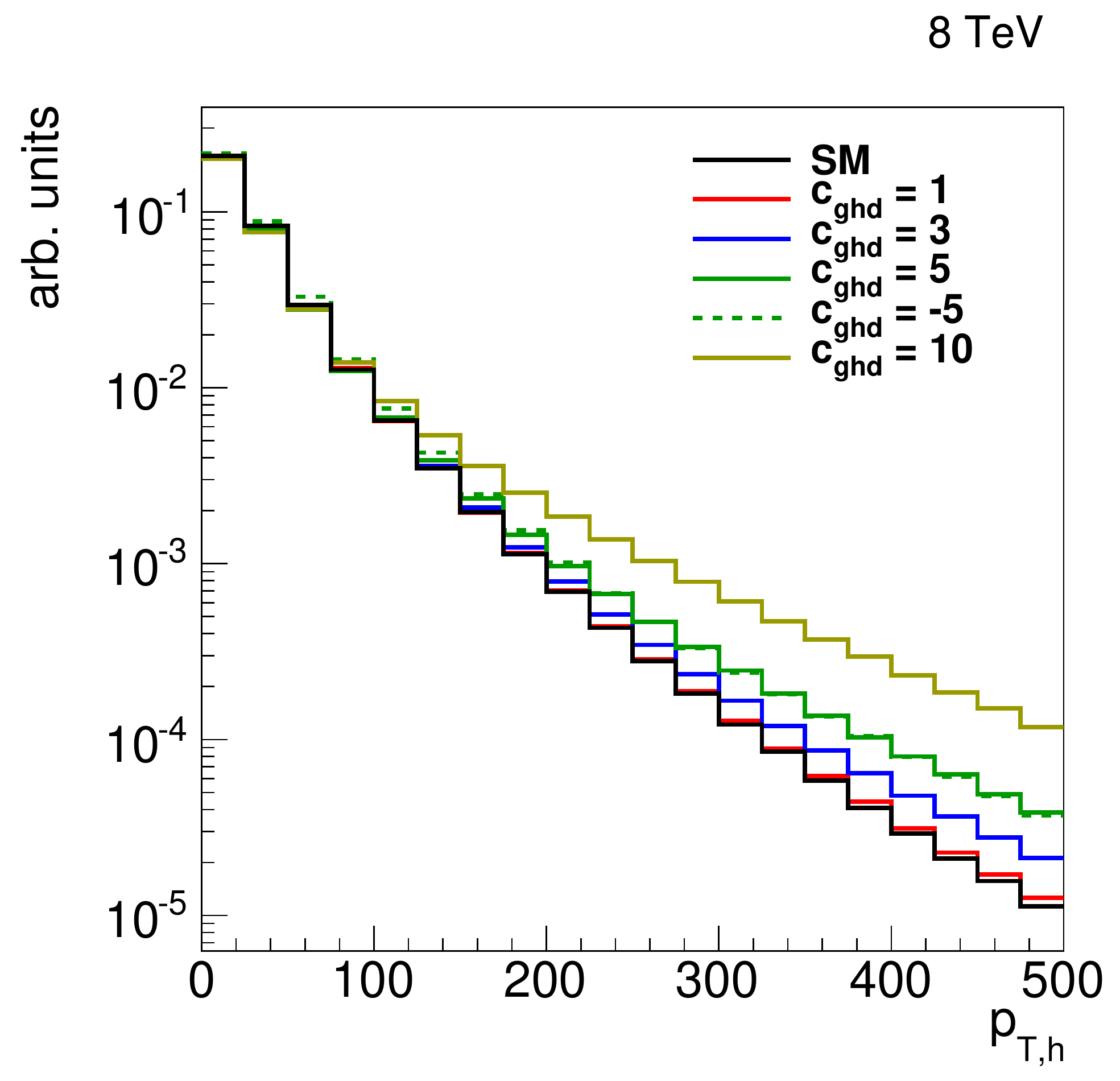}
\includegraphics[width=0.45\textwidth]{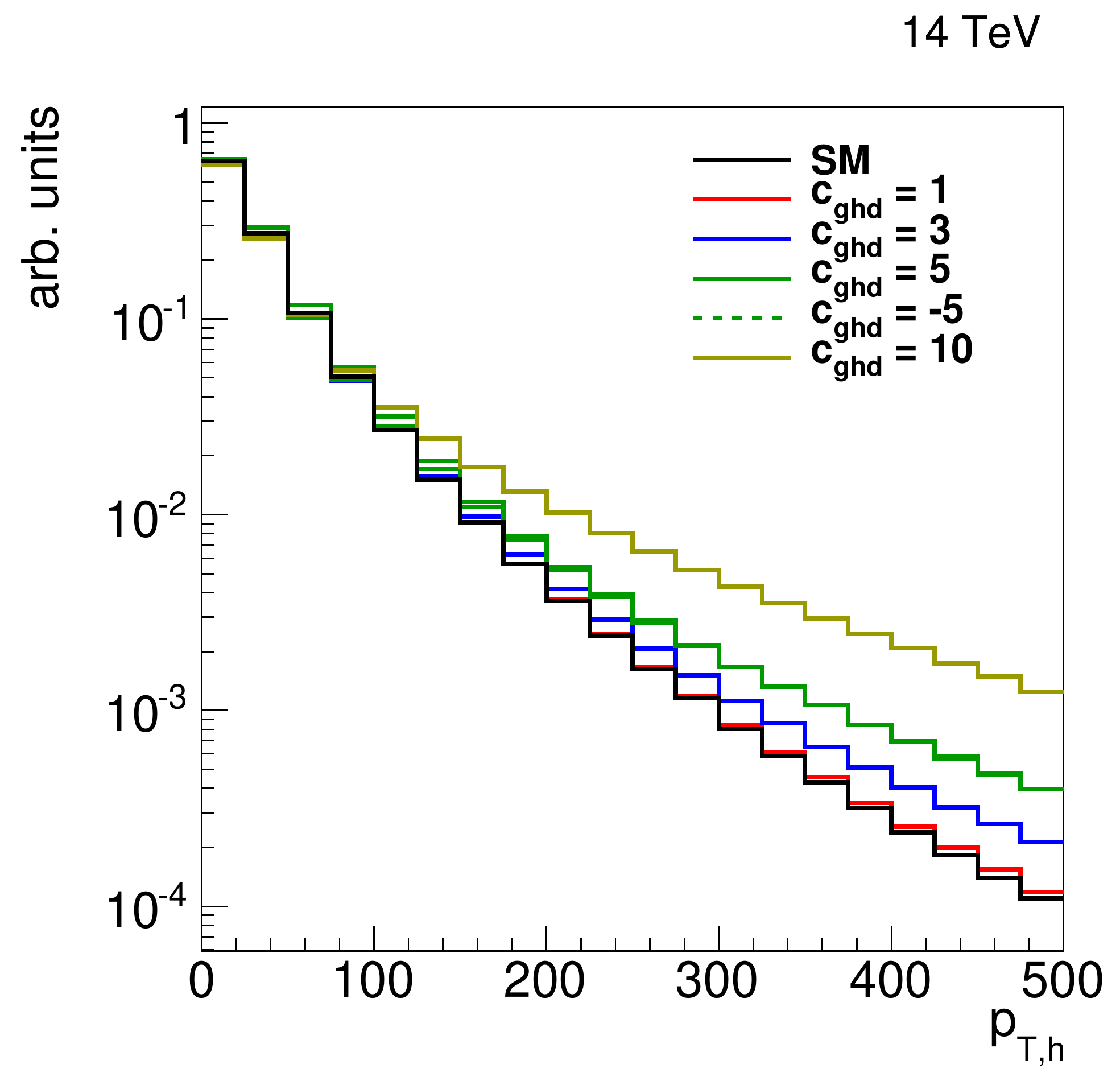}
\caption{The parton level
	transverse momentum of the Higgs boson in $b
	\overline{b} h$ final states (at leading order) is shown for five different values of the coupling
	$c_{ghd}$. The black, red, blue, dashed green, green, and yellow lines, listed from bottom to top, show the Higgs $p_T$ for $c_{ghd} = 0, 1, 3,
	+5, -5$, and $10$, respectively. These plots are made assuming an equal number of events
    for each distinct value of $c_{ghd}$, hence the kinematic differences
	arise independent of new physics alterations to the total cross-section
	for $pp \rightarrow b \overline{b} h$. The left plot shows the
	distributions for 8 TeV, and the right plot contains the 14 TeV
	distributions. Both sets of curves were generated with CTEQ6L parton distribution functions, default scale choices, and all parton-level cuts set to default MadGraph5 aMC@NLO values.}
\label{fig:higgspt}
\end{figure}

To quantitatively study how the gluon momentum dependence of $\mathcal O_{ghd}$
affects the kinematics in $b\bar b h$ events we implemented $\mathcal{O}_{ghd}$,
along with
standard model effective couplings of the Higgs boson to gluons and photons
\cite{Rizzo:1979mf,Kauffman:1996ix,Kauffman:1998yg}, in FeynRules
\cite{Alloul:2013bka,Alloul:2013naa} and generated parton level events using
MadGraph5 aMC@NLO \cite{Alwall:2014hca}. Throughout this paper, we will refer to
this model as ``SM + $\mathcal O_{ghd}$". The resulting Higgs $p_T$ spectra in $pp \to
b\bar b h$ events is shown in Fig.~\ref{fig:higgspt} for several different
values of $c_{ghd}$. Clearly, as $|c_{ghd}|$ is increased, there is a trend
towards more high-$p_T$ Higgs bosons, as expected from the Higgs momentum
dependence of the chromomagnetic dipole operator. At the same time, the spectra
from all $c_{ghd}$ and the SM coincide at low $p_{T,h}$. Depending on $c_{ghd}$,
the spectra begin to differ from the SM curve at $p_{T,h} \sim 100-200\,\gev$;
while these $p_{T,h}$ values are high, the center of mass energies they
correspond to are still small compared to $\Lambda$.

Another way to present the kinematic effect of $\mathcal O_{ghd}$ is via cumulative $p_T$ distributions, i.e. the number of Higgses with $p_T > p_{T,cut}$, divided by the corresponding number in the SM, as a function of $p_{T,cut}$. The cumulative distributions are shown below in Fig.~\ref{fig:ptsum} for the same set of $c_{ghd}$ as in Fig.~\ref{fig:higgspt}. The distributions show the same trend as Fig.~\ref{fig:higgspt}, though the fact that we have integrated over several $p_T$ bins makes the differences between smaller $c_{ghd}$ more evident. As one example, for a $p_{T,h}$ cut of $200\,\gev$, we expect approximately twice as many events in a $c_{ghd} = 5$, ``SM + $\mathcal O_{ghd}$" scenario than in the SM.

\begin{figure}[h!]
\includegraphics[width=0.45\textwidth]{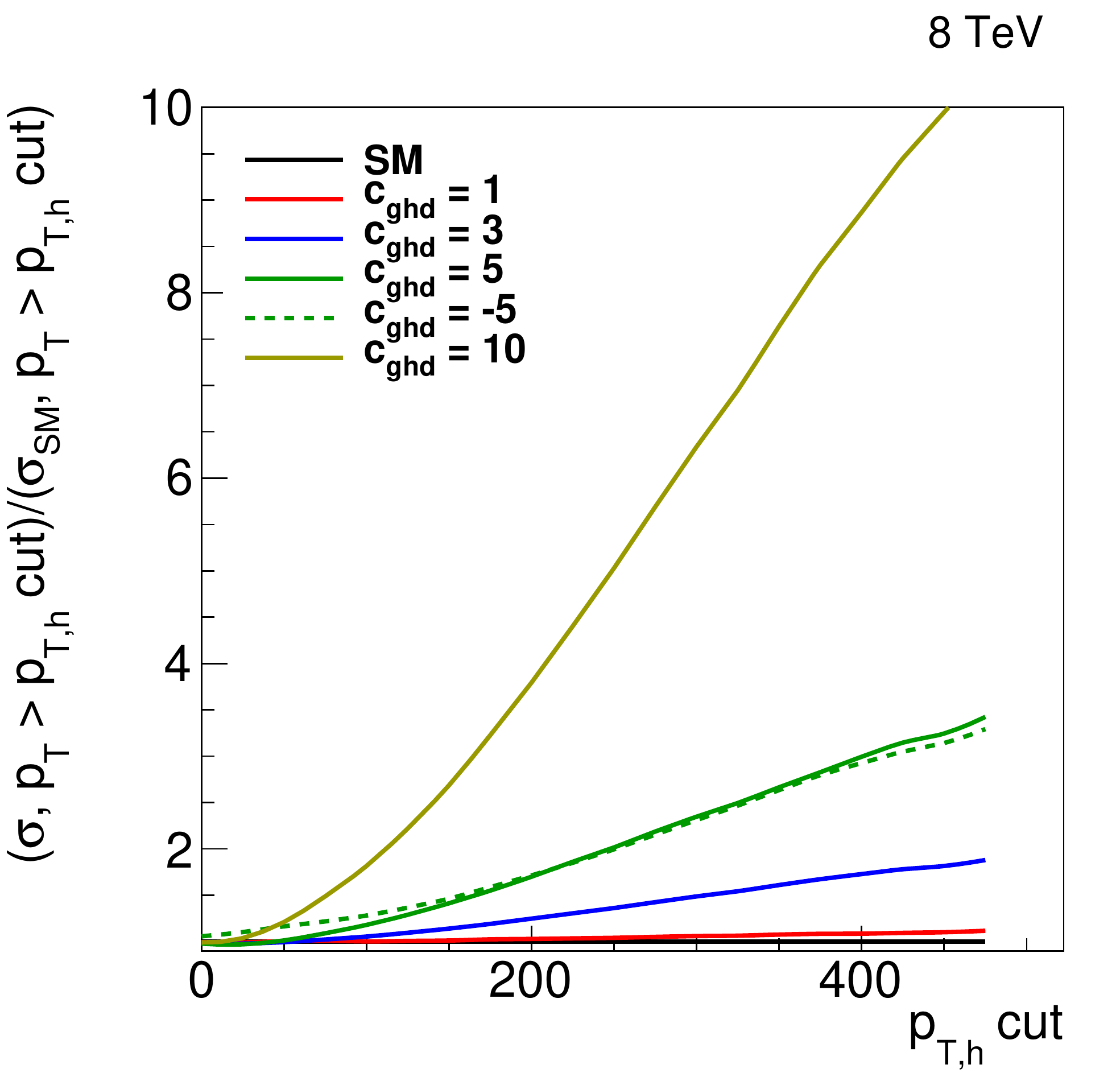}
\includegraphics[width=0.45\textwidth]{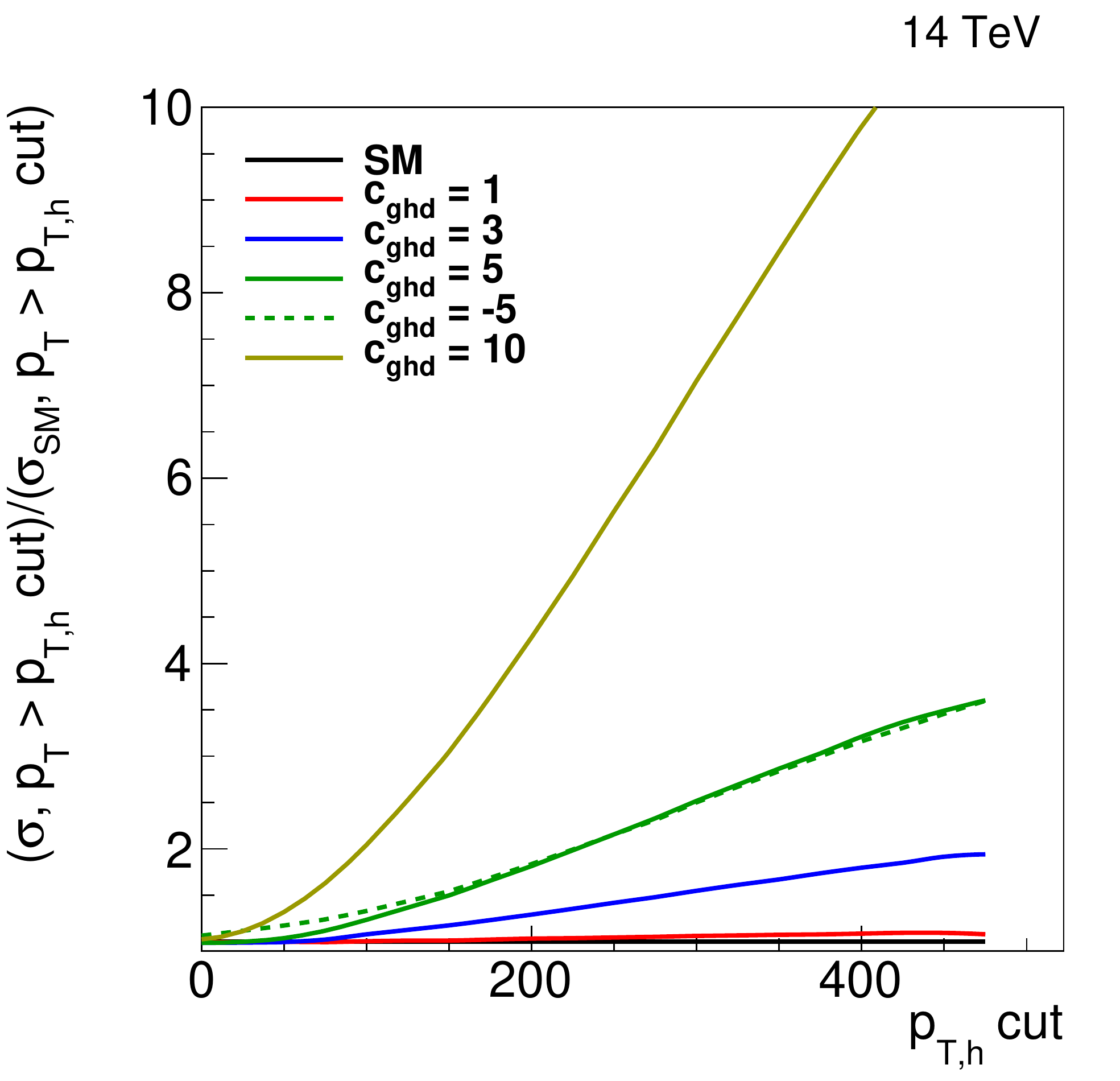}
\caption{The rate of parton-level, leading order $pp \to b\bar b h$ events with Higgs $p_T$ above a set cut value $p_{T, cut}$ as a function of $p_{T,cut}$. The number is normalized to the SM rate after applying the the same cut criteria. A number of different $c_{ghd}$ values are shown, using the same coloring scheme as Fig.~\ref{fig:higgspt} (from bottom to top, the curves are for $c_{ghd} = 0, 1, 3,
	+5, -5$, and $10$, respectively). }
\label{fig:ptsum}
\end{figure}

\subsection{Quark chirality and the chromomagnetic dipole}
\label{subsec:chiralandchrome}

In addition to dependence on the gluon momentum, another important structural
difference between
$\mathcal O_{ghd}$ and the SM is the chirality of the $b$ quarks. All of the
interactions generated from $\mathcal O_{ghd}$ involve a pair of bottom quarks
with opposite chirality ($b^{\dag}_L\,b_R$ or $b^{\dag}_R\,b_L$), while the SM
quark chirality depends on the interaction; for QCD gauge interactions, the
participating quarks have the same chirality, while for Higgs interactions the
quarks have opposite chirality\footnote{We are ignoring electroweak interactions
here as they are subdominant, and we only mention the Higgs interaction because
we are interested in final states containing Higgses.}. To see how the $b$-quark
chirality difference changes things, let us compare SM and SM + $\mathcal
O_{ghd}$ contributions to $pp \to b\bar b$. Starting with a $gg$ initial state
(though the same argument will work with $q\bar q$), we form $gg\to b\bar b$ by
sewing in either one triple-gluon interaction and one $b\bar bg$ interaction or
two $b\bar b g$ interactions. In either case, the two outgoing $b$ quarks have
the same chirality if we use only SM QCD interactions. Swapping out one of the SM
$b\bar bg$ vertices for the $\mathcal O_{ghd}$ $b\bar b g$ vertex, the outgoing
bottom quarks now have opposite chirality. In order for these two $gg \to \bar b
b$ contributions to interfere, the chirality on one of the bottom quark lines
needs to flip via a mass insertion. Thus, SM-$\mathcal O_{ghd}$ interference at
leading order (i.e. linear in $c_{ghd}$) in $pp \to \bar b b$ is suppressed by a
factor of $m_b$. 

The same line of logic, though slightly more complicated, holds for $pp \to \bar
b b h$. In the SM, the $\bar b b h$ final state arises by dressing the $pp \to
\bar b b$ diagrams discussed above with a Higgs emission on one of the $b$ quark
lines. The outgoing quarks therefore always have opposite chirality. At leading
order in $c_{ghd}$, the bottom quark chirality depends on whether the diagram
involves the $b\bar b g$ vertex contained in $\mathcal O_{ghd}$ or the
$b\bar b g h $ vertex. If the diagram contains a $\mathcal O_{ghd}$ $b\bar b g$ vertex, the outgoing $b$ have the same chirality because there are two chirality flips -- one in the $\mathcal O_{ghd}$ vertex and one in the subsequent Higgs emission (via a SM vertex). However, if the diagram contains a $\mathcal O_{ghd}$ $b\bar b g h$ vertex the $b$ and Higgs are emitted from the same vertex, thus there is only one chirality flip and the outgoing quarks have opposite chirality. 

The chirality structure of $pp \to \bar b b$ and $pp \to \bar b b h$ in the SM and $\mathcal O_{ghd}$ is summarized below in Table~\ref{tab:chirality}.  Any time the SM and $\mathcal O_{ghd}$ contributions have different chirality, interference between the two is suppressed by $m_b$.
\begin{table}[h!]
\begin{tabular}{|c|c|c|}
\hline Model & Process & Quark Chirality \\ \hline
SM & $pp \to \bar b b$ & S \\
SM & $pp \to \bar b b h$ & O \\ \hline
$\mathcal O_{ghd}(\bar b bg)$& $pp \to \bar b b$ & O \\ 
$\mathcal O_{ghd}(\bar b b g)$ & $pp \to \bar b b h$ & S \\
$\mathcal O_{ghd}(\bar b b g h)$ &$pp \to \bar b b h$ & O \\ \hline
\end{tabular}
\caption{Chirality of the outgoing bottom quarks in $pp \to \bar b b$ and $pp \to \bar b b h$ events within the SM and with leading order $O_{ghd}$ effects: S stands for same chirality, and O for opposite. The $\mathcal O_{ghd}$ vertex is indicated in parenthesis in the first column. }
\label{tab:chirality}
\end{table}

\subsection{Other operators and general EFT remarks}
\label{sec:otherops}

Having identified the structural aspects of $\mathcal O_{ghd}$ that lead to non-SM
kinematics, we can inspect other dimension-6 operators for similar features. The
dimension-6 operators that include $b$ quarks and Higgses are
\begin{equation}
\begin{aligned}
\mathcal O_{bHq} &= (Q^{\dag}_{3}\bar{\sigma}^{\mu}\,
Q_{3})(H^{\dag}\overleftrightarrow{D_{\mu}}H),
\quad \mathcal O_{c'Hd} = (
Q^{\dag}_{3}\tau_i \bar{\sigma}^{\mu}\,
Q_{3})(H^{\dag}\tau^i\overleftrightarrow{D_{\mu}}H), \\
\quad \mathcal O_{cHd} &=
(\bar b^{c\dag} \bar{\sigma}^{\mu}\, b^c)(H^{\dag}\overleftrightarrow{D_{\mu}}H),
\qquad \; \mathcal O_{y_d} = H^{\dag}H\, Q^{\dag}_{3} H\, b^{c\dag} + h.c.,
\end{aligned}
\end{equation}
where $\tau^i$ are the $SU(2)_w$ generators.
The first three operators always contain an electroweak gauge boson; they
correct the $\bar b b Z,\ \bar b t W$, etc.\ vertices and lead to 4-particle
$\bar b b W/Z h$ interactions. At $O(\alpha_s)$, these processes do not affect $pp
\to b\bar b$, $pp \to b\bar b h$, or $pp \to h$. In order to
accommodate electroweak precision constraints the couplings multiplying $\mathcal O_{bHq}$, 
$\mathcal O_{c'Hd}$, and $\mathcal O_{cHd}$ must be so small that they are unlikely to appreciably
alter LHC measurements. The fourth operator, $\mathcal O_{y_d}$, modifies the
relation between the $b$ mass and the Yukawa coupling, but it has the same
Yukawa structure as the SM. Thus the effects of $\mathcal O_{y_d}$ can be
subsumed into a rescaling of the SM rate. 

There are also dimension-six operators that contain only a subset of the fields
we are interested in -- $b\bar b$, Higgs, or gluon -- but which can affect both
the rate and kinematics of $b
\bar{b} h$ events. One such operator is
\begin{equation}
\mathcal{O}_{GH} = (H^{\dag}H )(G_{\mu \nu}^a G^{\mu \nu}_a ),
\end{equation}
which can feed into $p p \to b\bar b h$ as shown below in Fig.~\ref{fig:bbhhgg}, carrying non-SM momentum dependence. 

\begin{figure}[h!]
\begin{tabular}{c}
\includegraphics[width=2.in]{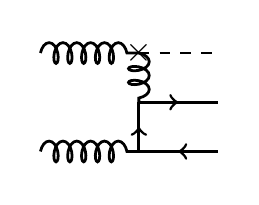}
\end{tabular}
\caption{This diagram shows one contribution of the Higgs-gluon kinetic coupling term, $\mathcal{O}_{GH}$, to $b \bar{b} h$ processes at the LHC.}
\label{fig:bbhhgg}
\end{figure}
However, as $\mathcal{O}_{GH}$ generates a tree-level contribution to $pp \to h$, its coefficient is
highly constrained by LHC Higgs data. It is possible to avoid these constraints
by including other operators in addition to $\mathcal{O}_{GH}$ and tuning the
strength of the multiple interactions against each other such that $pp \to h$ is
within the experimental limits. This sort of tuning is necessary when looking
for non-SM kinematic features is $pp \to t\bar t h$
events~\cite{Degrande:2012gr,Bramante:2014gda}. Taken individually, the
couplings of $\mathcal{O}_{GH}$ and the top quark chromomagnetic dipole are
constrained to be small. However if both operators are turned on simultaneously
they can be arranged to cancel, admitting  larger couplings -- and thereby
larger effects in Higgs kinematics -- without running afoul of bounds. As we
will see shortly, the constraints on
$\mathcal{O}_{ghd}$ from inclusive $pp \to h$ production are small, hence we are
less restrained by phenomenological restrictions, and can omit
$\mathcal{O}_{GH}$ when exploring non-SM Higgs kinematics in $b\bar b h$ events.

In this study we will focus solely on the
chromomagnetic $b$-quark operator $\mathcal{O}_{ghd}$. This means that our analysis
neglects any higher order (dimension 6 and above) operators that may only include gluons, bottoms quarks or
the Higgs. A rationale for this approach is as follows: without a UV completion,
evaluating the effect of a truncated set of cutoff-suppressed
operators is a delicate and potentially ambiguous task. The chromomagnetic dipole is the
lowest dimension operator that gives a non-SM kinematic morphology to $b \bar b h$ events. So while other $\rm{dim} \geq 6$ operators may change the rate of $b \bar b h$, a search for $\mathcal{O}_{ghd}$ will be sensitive to both these changes and shifts in final state particle kinematics.

One has to pay special attention to the choice of the
cutoff suppressing the $\mathcal O_{ghd}$ operator. For a given effective
cutoff, absorbing all powers of couplings into the scale $\Lambda_{\eff}$ means that effects from neglected $d > 6$ terms are naively suppressed relative to lower order terms by
powers of $\sqrt{\hat s}/\Lambda_{\eff}$. There could be some subtleties regarding
unnaturally small couplings or operators generated by loop diagrams that are
impossible to address in an effective field theory without the benefit of a UV
completion. Our dilemma is similar to that faced by mono-jet dark matter collider searches (e.g. Ref.~\cite{Birkedal:2004xn,Beltran:2010ww,Goodman:2010ku,Bai:2010hh}), where the cuts imposed on the initial state jet are often
close to the scale suppressing the SM-DM interactions.

To illustrate, note that as we increase the coefficient $c_{ghd}$ for fixed $\Lambda$, the
effective cutoff $\Lambda_{\eff}$ decreases. For smaller values of $c_{ghd}$,
$\sqrt{\hat s}/\Lambda_{\eff}$ will always be much less than one for LHC energies, but
for larger values of $c_{ghd}$ one can start getting into a situation  where the
LHC has probed $\sqrt{\hat s} >\Lambda_{\eff}$, but not necessarily in the process we
are bounding. The highest scale probed so far in a process affected by
$\mathcal{O}_{ghd}$  ($pp \to b\bar b$) is $\sim 3\,\tev$. When bounding
$\mathcal{O}_{ghd}$, we will encounter processes that can only be seen in current or future LHC data when
$\Lambda_{\eff} < 6\,\tev$.
We quote these bounds with the caution that they are
more a statement of what it takes to deviate from the SM and what sort of UV
physics is allowed rather than an actual coupling value. For example, quoting a bound
$\Lambda_{\eff} \sim 1\,\tev$ requires that whatever UV completion kicks in at
$1\,\tev$ not disrupt high-mass $pp \to  b \bar b$ studies. In order
to remain agnostic towards the UV completion of the theory, we will consider cutoffs
as small as $\Lambda_{\eff} \gtrsim 1\,\tev$, although as just explained, cutoffs
smaller than $\sim 3\,\tev$ should be considered heuristic, because e.g. any
resonant $s$-channel production of $b \bar b$ at this energy would likely have
been seen in $b \bar b$ dijet studies. For a detailed discussion of the delicate
issues surrounding the task of bounding effective field theories with data from
colliders, see Ref.~\cite{Englert:2014cva}.

\section{Constraints on the bottom quark chromomagnetic dipole}
\label{sec:constraints}

Because devoted SM Higgs plus jets studies including $b$-tags have not yet been conducted at the LHC, 
the chromomagnetic dipole operator $\mathcal{O}_{ghd}$ remains unconstrained
from direct $pp \rightarrow b \bar b h$ measurements. However, a number of other
processes affected by a $b$-quark chromomagnetic dipole can put an upper bound
on $c_{ghd}$. 

Before determining non-$b \bar b h$ LHC particle
production constraints on $\mathcal{O}_{ghd}$, we comment on how $\mathcal
O_{ghd}$ affects static Higgs properties. Specifically, $\mathcal O_{ghd}$
contributes to the three-body Higgs decay mode $h \to b \bar b g$. In principle
this changes the Higgs width, which affects all Higgs branching ratios. In
practice we find this partial width is small, less than $1/1000$ of the total
Higgs width for $O(1)$ values of $c_{ghd}$, so that the total Higgs width is
essentially independent of $c_{ghd}$ for the range of values considered in this
study. Note that Higgs width and branching ratio constraints are more severe on other
possible higher-dimension Higgs-bottom quark interactions which contribute
directly to $h\to b\bar b$, such as $\mathcal
O_{yb} = (H^{\dag}H)(Q^{\dag}H)\,y_b\, b^c$.

\subsection{$b$-jet production; $pp (\bar p) \to b \bar b$}
\label{subsec:bbprod}

In Ref.~\cite{Bramante:2014gda}, it was shown that measurements of the inclusive
$p p \rightarrow t \bar{t}$ cross-section
set strong constraints on the size of the top quark chromomagnetic dipole, the
up-type quark cousin of the operator in Eq.~(\ref{eq:bchrome}):
\begin{equation}
\mathcal O_{ght} \sim  \frac{c_{ght}}{\Lambda^2_0}\,(Q_3 H)\,y_t\,\sigma_{\mu\nu}t^{A}\, t^c\, G_A^{\mu\nu}.
\label{eq:tchrome}
\end{equation}
Using the same convention as in Eq.~(\ref{eq:masscouplconv}), we can combine the Yukawa coupling and original cutoff $\Lambda_0$ into a new cutoff, $y_t/\Lambda^2_0 = 1/\Lambda^2$. However, as the top Yukawa is nearly $1$, $\Lambda$ and $\Lambda_0$ are approximately equal.

For a cutoff scale $\Lambda_0 =1\, \tev$, the LHC $pp\to t\bar t$ measurements
\cite{Aad:2012vip,Chatrchyan:2013faa,Aad:2014kva,Khachatryan:2014ewa} restrict
$c_{ght}$ to $-1 \lesssim
c_{ght}\, \lesssim 0.5$. Based on this observation, one might
expect measurements of $pp \rightarrow b \bar{b}$ to have a similar impact on
the size of $c_{ghd}$, the coefficient of $\mathcal O_{ghd}$. However, this
expectation is not correct for a couple of reasons. First, the SM cross section
$pp \to b\bar b$ is orders of magnitude larger than $pp \to t\bar t$ and is
dominated by low energy scattering at $\hat s \sim 4\,m^2_b$. The effect of
$\mathcal O_{ghd}$ is suppressed at these low energies, making the inclusive $pp
\to b\bar b$ cross section a useless observable for bounding $c_{ghd}$. To have
any chance at sensitivity to $\mathcal O_{ghd}$, we must focus on the most
energetic $pp \to b\bar b$ collisions. To date, the best measurement of $b
\bar{b}$ production at large center-of-mass
energies has been set by a CMS study excluding heavy $b \bar{b}$ resonances with
invariant mass ranging from $\sim 1-5 {~ \rm TeV}$ \cite{CMS-PAS-EXO-12-023}.

A second reason the lesson from $pp \to t\bar t$ does not carry over to $pp \to
b\bar b$ is interference. As mention in Sec.~\ref{subsec:chiralandchrome}, the
chromomagnetic moment operators lead to gluon-quark-quark interactions with a
different chirality structure than the usual SM vertex, thus interference between
new physics and the SM must be proportional to the quark mass. As a result,
interference in $pp \to b\bar b$ is highly suppressed, while in $pp \to t\bar t$
it is not.

 Finally, when comparing $\mathcal O_{ghd}$ and $\mathcal O_{hgt}$, we must remember our cutoff convention. While the convention choice does not change the results, when comparing different operators, we have to make sure we use the same rules. As $y_t \sim 1$ there is no difference between the initial cutoff $\Lambda_0$ and the rescaled cutoff $\Lambda$, while the small value of $y_b$ leads to $\Lambda \sim 6\,\Lambda_0$. Stated another way, a coefficient $c_{ght} = 1$ leads to the same overall operator strength as $c_{ghd} = 40$. 

\begin{figure}[h!]
\begin{tabular}{c}
\includegraphics[scale=.7]{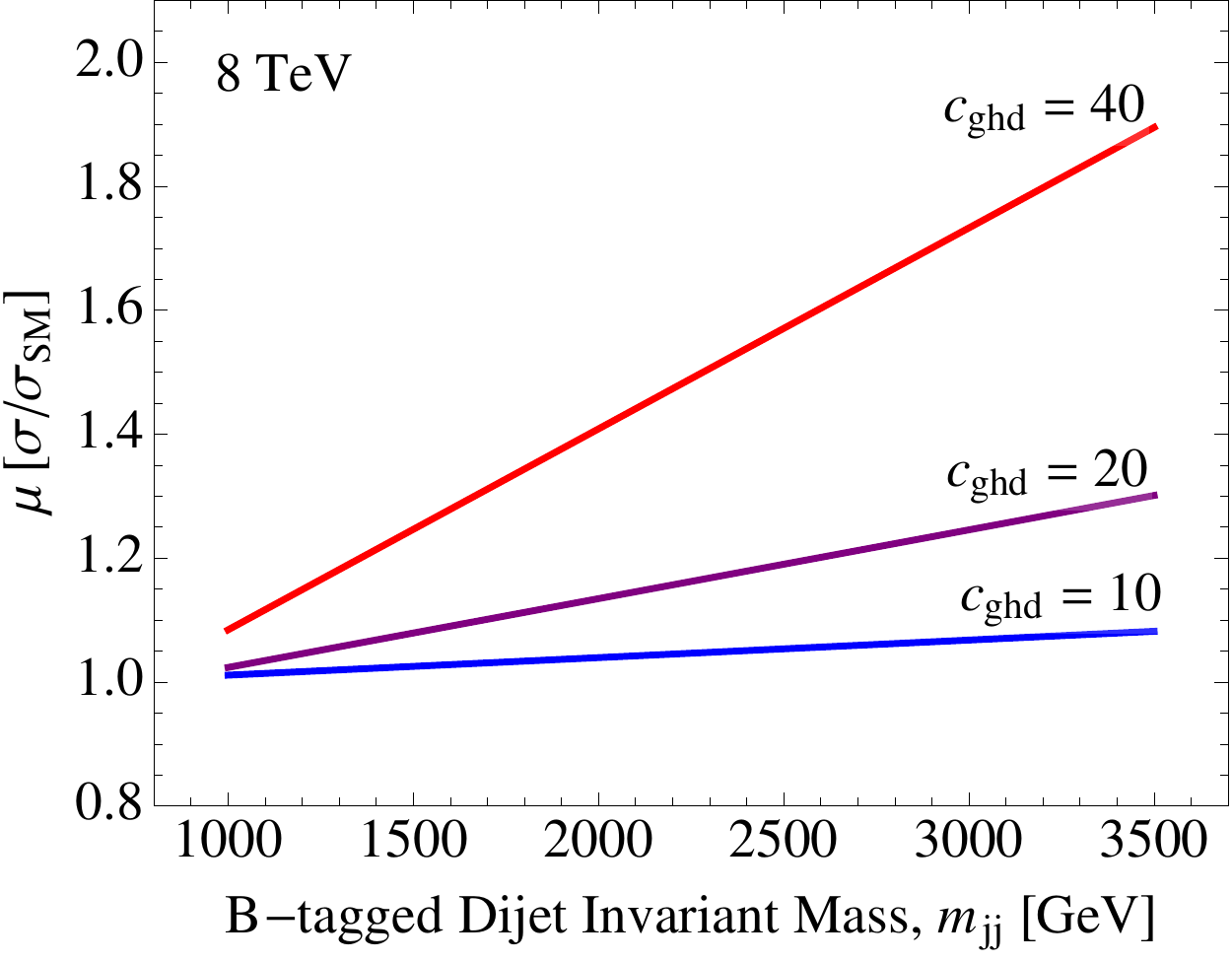}
\end{tabular}
\caption{This plot shows the ratio of new physics to Standard Model $pp \rightarrow b \bar b$ cross-sections ($\mu^{\rm{parton}} = \frac{\sigma}{ \sigma_{SM}}$) plotted against the $b$ quarks' dijet invariant mass, for chromomagnetic dipole couplings ($c_{ghd}$) as indicated, and for $\sqrt{s} = 8~ \tev$ proton collisions. Additional cuts applied included requiring $2$ or more jets with $p_T > 30\,\gev$), with $|\eta_j| < 2.5$, and $m_{jj} > 1\,\tev$. Measurements of the area normalized cross-section vs. $b$-tagged dijet invariant mass \cite{CMS-PAS-EXO-12-023} indicate that $c_{ghd}=40$ is ruled out, while $c_{ghd}=20,10$ are allowed at $\sim 95\%$ confidence.}
\label{fig:mudijet}
\end{figure}

To determine the sensitivity of high-mass $pp \to b\bar b$ searches to the
bottom quark chromomagnetic operator, we mimicked the analysis of
Ref.~\cite{CMS-PAS-EXO-12-023}. Specifically, we generated $pp \to b\bar b$
events at $\sqrt{s} = 8~ \tev$ using the ``SM + $\mathcal O_{ghd}$" Madgraph
model introduced in Sec.~\ref{sec:bboost} for several different $c_{ghd}$
values. These parton level events were subjected to the following cuts: $2$ or
more jets (anti-$k_T$, jet radius $0.5$, and $p_T > 30\,\gev$), with the leading
two jets satisfying $|\eta_j| < 2.5$, $\Delta \eta_{jj} < 1.3$ and $m_{jj} >
1\,\tev$. We plot the relative SM and ``SM + $\mathcal O_{ghd}$" rate per dijet
invariant mass bin in Fig.~\ref{fig:mudijet}. We do not simulate parton
showering and detector acceptance, since these will affect SM and new physics $b
\bar b$ equally.

Because the LHC measurement
\cite{CMS-PAS-EXO-12-023} normalizes dijet mass distributions to the
shape of SM Monte Carlo distributions and is not an absolute cross-section measurement, it is
insensitive to overall shifts in $b\bar{b}$ rates. For a very large value of
$c_{ghd} = 40$ (meaning a cutoff $\Lambda = 1~{\rm TeV}$), we find that current
$pp \to b\bar b $ are barely sensitive to $\mathcal O_{ghd}$, and this affect
would require careful comparison of event rates in low and high invariant mass
bins. Specifically, for $c_{ghd} = 40$, we find the rate difference between the
SM and the SM augmented by $\mathcal O_{ghd}$ in $1\,\tev \le m_{bb} \le
5\,\tev$ events is nearly double, which is excludable using the dijet invariant
mass event distributions in \cite{CMS-PAS-EXO-12-023}. The dijet invariant mass
event bins of \cite{CMS-PAS-EXO-12-023} have $1 \sigma$ error bars of $\sim
20\%$, so a bin-to-bin shift in expected event rate must be roughly double for a
$5 \sigma$ exclusion. This qualitative comparison was carried out only at the
parton level. We do not expect showering, hadronization, and detector effects to
significantly change the story, especially as the event rate is fitted to the
data and no detailed event counts or error bars are given
in~\cite{CMS-PAS-EXO-12-023}.

A final complication in these $pp \to \bar b b$ limits on $\mathcal O_{ghd}$ is
how we model the proton, i.e.~the parton distribution functions. The bottom
quark is light enough that it can compose a non-negligible fraction of
sufficiently high-energy protons. Admitting initial state $b/\bar b$ quarks
leads to new contributions to $pp \to \bar b b$, thereby introducing new ways
$\mathcal O_{ghd}$ can enter. However, we find that including $b$ quarks as
initial state partons enhanced $pp \rightarrow b \bar{b}$ rates for the SM and
$c_{ghd} = 40$ equally, in each case increasing the cross-section by $\sim 3\%$
in each $b \bar{b}$ invariant mass bin (CTEQ6L parton distribution functions).
Since the normalization shift is the same with and without $\mathcal O_{ghd}$,
the addition of $b$ quark partonic states does not alter the constraints on
$c_{ghd}$ presented  in Figure \ref{fig:mudijet}.

We conclude that current studies of high invariant mass $b \bar{b}$ production
are not sensitive to the $b$ quark chromomagnetic dipole for couplings less than $c_{ghd}=40$. As we have discussed, this is partly because the interference of the $b$ quark chromomagnetic operator's $pp \rightarrow b
\bar{b}$ production and SM $b \bar{b}$ production requires a
chirality flip for one of the $b$ quarks, and a corresponding suppression from the
bottom quark mass, as illustrated in Fig. \ref{fig:bbhdiag}.

\subsection{Inclusive Higgs production: $pp \to h$}
\label{subsec:higgsprod}

Having shown that $pp \to b\bar b$ events do not place any strong constraint on the chromomagnetic bottom quark operator, we now explore constraints from SM Higgs production at 7 and 8 TeV. Because $\mathcal O_{ghd}$ is generated
at the scale $\Lambda$, if we evolve down to the scale relevant to Higgs
production via the renormalization group, the loop-level effects of $\mathcal O_{ghd}$ can be subsumed into an
altered coefficient for the effective $h\,G^{\mu\nu}_A\,G_A^{\mu\nu}$ operator: 
\begin{align}
\mathcal L \supset c^{SM}_{hgg} h\,G^{\mu\nu}_A\,G_A^{\mu\nu} \to \Big(c^{SM}_{hgg} + \frac{c_{ghd}\,m_b}{16\pi^2\,\Lambda^2}\log\Big(\frac{\Lambda}{m_b}\Big) \Big)\,h\,G^{\mu\nu}_A\,G_A^{\mu\nu},
\label{eq:higgsloopcorr}
\end{align}
where $c^{SM}_{gg}$ is the usual SM contribution coming from top quark loops.

One might expect that the above alteration of Higgs boson production would be
the most constraining result of a bottom quark chromomagnetic dipole, especially
since prior studies of the top quark chromomagnetic dipole found this to be the
case for the top quark \cite{Degrande:2012gr,Bramante:2014gda}.

\begin{figure}[h!]
\begin{tabular}{c}
\includegraphics[scale=1.5]{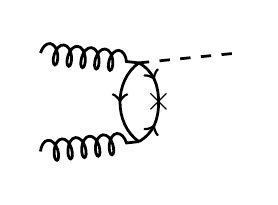}
\end{tabular}
\caption{This diagram shows one contribution of the bottom quark
	chromo-dipole operator (\ref{eq:ghd}) to Higgs boson
	production at the LHC. Note that the chirality flip in the quark loop will
	introduce a suppression proportional to the bottom Yukawa $y_b$ for the case of
	a $b$ quark chromomagnetic dipole contributing to Higgs boson production.}
\label{fig:higgsloop}
\end{figure}
 
However, the $b$-quark lines emanating from $\mathcal O_{ghd}$ have opposite
chirality. Therefore, following the same logic as in
Sec.~\ref{subsec:chiralandchrome}, in order to close the $b$-quark lines from
$\mathcal O_{ghd}$ into a loop
contribution to $gg \to h$, a mass insertion on one of the internal quark lines
is required. The chirality structure for a sample diagram is shown explicitly in
Fig. \ref{fig:higgsloop}. This mass insertion causes
the extra factor of $m_b$ in the second term in Eq.~(\ref{eq:higgsloopcorr}) which, 
since $m_b$ is much less than the other scales ($\Lambda$ or $v$), strongly
suppresses the $\mathcal O_{ghd}$ contribution. An
analogous chirality flip is required to generate a $gg\to h$ contribution from
the top-quark chromomagnetic operator, as done in
Ref.~\cite{Degrande:2012gr,Bramante:2014gda}. However, in that case the large
mass of the top quark makes the mass-insertion suppression price much less
severe.

Plugging representative numbers into Eq.~(\ref{eq:higgsloopcorr}), we see that for a cutoff of $\Lambda = 6 ~{\rm TeV}$, the change\footnote{We use $c_{hgg} \simeq
\frac{\alpha_s(m_Z)}{3 \pi v}\left(1+ \frac{7}{30} x + \frac{2}{21} x^2\right)$
where $x \equiv \frac{m_h^2}{4m_t^2}$.} in $c_{hgg}$ is very small: $\delta c_{hgg} \sim 10^{-4}$ for $c_{ghd} = 1$. For larger $c_{ghd}$,  $\delta c_{hgg}$ increases, but only linearly with $c_{ghd}$. In order to have a noticeable effect on inclusive Higgs production, the coupling would need to be $O(100)$, an unreasonably large value from the EFT perspective. We conclude that inclusive
Higgs production is not a sensitive probe of the bottom quark chromomagnetic
dipole operator. This should be contrasted with the top-quark chromomagnetic dipole:
there corrections to $c_{hgg}$ from the dipole contribution are so large that
$c_{ght}$ is constrained -- for $\Lambda = \tev$ and assuming no other
higher-dimensional operators -- to $\lesssim O(1)$. 

\subsection{Higgs plus dijets: $pp \to h+ j j$}
\label{subsec:jjh}

We now address a final state at the LHC which has the potential to constrain
$\mathcal{O}_{ghd}$, $pp \to h + jj$. While there is no dedicated $\bar b b + h$ search, $\bar b
bh$ events will show up in any Higgs plus jets search provided bottom quarks are
not explicitly vetoed. The $pp\to b\bar b h$ final state looks promising since,
as we saw in Sec.~\ref{sec:bboost} and \ref{subsec:chiralandchrome}, the
SM-$\mathcal O_{ghd}$ interference is not suppressed by $m_b$, and it is
possible to enhance the $\mathcal O_{ghd}$ effects by looking at regions of
phase space where $\sqrt{\hat s}$ is large. 

The Higgs plus jets channel with the tightest constraint on new physics
contributions is the diphoton decay mode, $h(\gamma\gamma)+jj$; specifically,
the rate in this channel in any extension of the SM relative to the SM -- the
signal strength $\mu^{\text{collider}} $ -- is restricted to $0.8\pm 0.7$ by
ATLAS~\cite{Aad:2014eha}\footnote{ATLAS categorizes diphoton Higgs events by
their production mechanism rather than the final state. The number quoted is the
vector boson fusion (VBF) category. While VBF events should comprise a large
fraction of $h + jj$ events, the two rates are not  equal. For this reason, we
use the CMS numbers and procedure throughout this section. }
, and $\mu^{\text{collider}}_{hjj} =
	1.11^{+0.32}_{-0.30}$ by CMS~\cite{Khachatryan:2014ira}. Since $\mathcal O_{ghd}$ has a different structure than the SM, we do not expect SM and ``SM + $\mathcal O_{ghd}$" events to have the same cut acceptance. Therefore, in order to see how the signal strength limits translate into bounds on $c_{ghd}$, we have to rely on Monte Carlo simulation. Specifically, we
simulate the CMS analysis of $pp \rightarrow h + jj \rightarrow \gamma \gamma +
jj$ final states \cite{Khachatryan:2014ira} for a number of values of $c_{ghd}$
and a cutoff of $\Lambda = 6 ~ {\rm TeV}$, then compare with the SM prediction. The signal  events are generated at parton level using the Madgraph ``SM + $\mathcal O_{ghd}$'' model, passed through Pythia 8.1
\cite{Sjostrand:2007gs} for showering and hadronization, then routed through Delphes \cite{deFavereau:2013fsa} to incorporate detector effects\footnote{We use the default Delphes CMS detector card in all analyses.}. 
We then apply the following cuts, taken from the CMS analysis~\cite{Khachatryan:2014ira}:
\begin{itemize}
\item[(1)] The highest $p_T$ photon must be larger than half the invariant mass of the photon pair, $p_{T,\gamma_1} > m_{\gamma \gamma}/2$, and the second highest $p_T$ photon must satisfy $p_{T,\gamma_2} > 25\,\gev$.
\item[(2)] The diphoton invariant mass, $m_{\gamma \gamma}$, must be between 120 GeV
	and 130 GeV.
\item[(3)] The two leading jets must have $|\eta|<4.7$.
\item[(4)] The difference in azimuthal angle between the dijet and diphoton systems ($\Delta \phi_{jj-\gamma\gamma}$) must be greater than 2.6.
\item[(5)] The Zeppenfeld variable $Z=\eta[\gamma_1 + \gamma_2] -
	(\eta[j_1]+\eta[j_2])/2$ must be less than 2.5.
\item[(6)] The difference in pseudorapidity between the jets ($\Delta \eta_{jj}$) must be greater than 3.
\item[(7)] For the dijet tight (dijet loose) cut the invariant mass of the jets must be $> 500 ~{\rm GeV}$ ($> 250 ~{\rm GeV}$) and both jets must have $p_T > 30 ~ {\rm GeV}$ (the second highest $p_T$ jet must have $p_T > 20 ~ {\rm GeV}$).

\end{itemize}  

Post-cuts, we compare the rate relative to the Standard Model expectation
(obtained by simulating SM events without the addition of a bottom quark
chromomagnetic dipole), and these ratios of rates for several $c_{ghd}$ are
shown in Table \ref{tab:hjj}. 
\begin{table}[h!]
{\renewcommand{\arraystretch}{1.4}
\begin{tabular}{|c||c||c||c|c|c|c|c|}
\hline
&Measured (\cite{Khachatryan:2014ira}) & $c_{ghd}$ = & -40 & -20 & 0 & 20 & 40
\\
\hline
$\mu_{hjj}$ & $1.11^{+0.32}_{-0.30}$ & & 1.19 & 1.03 & 1 & 1.04 & 1.24\\
\hline

\end{tabular}
}
\caption{This table shows the expected rate parameter $\mu^{\text{collider}}_{hjj}$ in a CMS study of $pp \rightarrow h + jj \rightarrow \gamma
	\gamma + jj$ for indicated values of a chromomagnetic bottom quark coupling
	$c_{ghd}$ with a cutoff of $\Lambda = 6~{\rm TeV}$. The signal strength $\mu^{\text{collider}}_{hjj}$ is calculated after collider-level cuts have been applied. 
	The signal events were
	generated via MadGraph 5, Pythia, and Delphes at $\sqrt{s} = 8 ~{\rm
	TeV}$ as described in the text. Cuts were implemented to match those of
	\cite{Khachatryan:2014ira}, and the rate from the CMS study $\mu^{\text{collider}}_{hjj} =
	1.11^{+0.32}_{-0.30}$ is displayed for reference.}
\label{tab:hjj}
\end{table}
We find that for a cutoff of $\Lambda = 6~{\rm TeV}$, even
couplings as large as $|c_{ghd}| \sim 40$ are not constrained by the most recent
Higgs plus jets studies.

Of course, a coupling this large and $\Lambda = 6~{\rm TeV}$ corresponds to the
same effective cutoff as an order one coupling and $\Lambda =1\, {\rm TeV}$.
Depending on the UV model inducing the chromomagnetic dipole, order one
couplings for such a $1\, \tev$ cutoff would be at odds with $b \bar{b}$ di-jet
resonance studies, which do not show any deviation from a Standard Model
invariant mass distribution. We discussed the bound on 1-6 TeV di-b-jet
invariant mass distributions in Section
\ref{subsec:bbprod} and found $pp \to \bar b b$ insensitive to $\mathcal
O_{gdb}$, however the bounds derived there were under the assumption that the
{\em only} new physics effect was the chromomagnetic moment. If the cutoff
$\Lambda_{\text{eff}}$ of the EFT is within the energy reach of $pp \to \bar b b$, we
must include all the physics at $\Lambda_{\text{eff}}$ -- either in the form of
on-shell states or dimension $> 6$ operators -- in order to get a meaningful
bound. Stated another way, one cannot simply take $c_{ghd} \sim 40, \Lambda =
6\,\tev$ as an indication of the scale of some new physics without explaining
via UV model building why experiments involving $b,\bar b, g$ or Higgs that are
sensitive to $\sqrt s > \Lambda_{\text{eff}}$ show no deviation.

Regardless of the UV physics being probed, it is clear from Table \ref{tab:hjj} that
flavor-blind $hjj$ studies are insensitive to the $b$ quark chromomagnetic
dipole.
 
\section{Press the $b$-tag button}
\label{sec:btagit}

Having seen that existing flavor blind $h + jj$ studies place no bound on $\mathcal{O}_{ghd}$, we now explore how requiring $b$-tags and modifying the cuts can improve the sensitivity. We first build off of the 8 TeV LHC
analysis discussed in the previous section~\cite{Khachatryan:2014ira}, then present a more
optimized approach for use at a high luminosity run of the 14 TeV LHC. For the
relatively low luminosity collected at 8 TeV, the cuts applied will need to be parsimonious to obtain substantial exclusion
significance.

\subsection{Constraints on $b \bar{b} h$ at the 8 TeV LHC}

 In addition to cuts (1)-(3) given in Subsection \ref{subsec:jjh},
for a $b \bar b h$ study at 8 TeV we also require that
\begin{itemize}
\item The event must have one jet that passes the medium selection
	criteria detailed in \cite{Chatrchyan:2012jua} (70\% b-jet tagging, 1.5\%
	light quark miss-tagging).
	\item The $p_T$  of the vector sum of the diphoton momenta must be greater than $150$ GeV.
\end{itemize}
For this study, we modify the default CMS Delphes card to tag $70\%$ of the
$b$-jets in our simulated events, and miss-tag $1.5\%$ of the light quarks as
$b$-jets.  While we retain cuts (1)-(3) from the CMS analysis, the remaining
four are dropped. These latter cuts, in particular the cut on the Zeppenfeld
variable~\cite{Rainwater:1996ud}, are designed to pick out Higgs events produced
via vector boson fusion (VBF). While separating Higgs plus jets events into
``gg-initiated'' and ``VBF" categories is an extremely useful tool for pinning
down exactly how the Higgs couples to vector bosons vs. gluons, these cuts do
more harm than good for our $b\bar b h$ signal. There are several $pp \to b\bar
b h$ diagrams that have similar topology to VBF-initiated Higgses, however the
feature of $\mathcal O_{ghd}$ we really want to exploit is the momentum
dependence. We find better results by dropping these VBF-specific cuts in favor
of the cut directly on the diphoton (Higgs) $p_T$.

As we have modified the analysis, we can no longer use quoted CMS backgrounds. In order to simulate background events arising from non-Higgs $j j \gamma
\gamma$ final states, and to check our collider simulations against the CMS
result, we produced a background sample of $pp \rightarrow j j \gamma \gamma$
and $pp \rightarrow b \bar{b} \gamma \gamma$ events in Madgraph. We then
followed the same procedures as previously outlined for showering and detector simulation of
this background\footnote{We checked the MadGraph-generated background sample
	of $j j \gamma \gamma$ events by using the same cuts implemented in
	\cite{Khachatryan:2014ira}, and found that for these cuts, the expected
numbers of background events matched those found by the CMS study to within
20\%.}.  

The results from this modified, $b$-tag enhanced analysis are listed in Table
\ref{tab:hbb} for a sampling of $c_{ghd}$ values. We quote event counts for
$20\,\fbinv$ of integrated luminosity. 
\begin{table}[h!]
	\renewcommand{\arraystretch}{1.4}
	\begin{tabular}{|c||c||c||c|c|c|c|c|c|}
		\hline
		& $ j j \gamma \gamma$ bkgd & $b \bar{b} \gamma \gamma$ bkgd & $c_{ghd} =$ & -40 & -20 & 0 & 20 & 40 \\
		\hline
		8 TeV $b \bar{b} h$ events (20 $\fbinv$) & 0.2 & 0.4 & & 1.9 & 0.5 & 0.05 & 0.5 & 1.8 \\
		\hline
		8 TeV $\mu_{b \bar{b} h}^{\text{collider}}$  & -- & -- & & 38 & 10 & 1 & 10 & 36 \\
		\hline
		8 TeV $\mu_{b \bar{b} h}^{\text{parton}}$  & -- & -- & & 2.7 & 1.5 & 1 & 1.2 & 2.1 \\
		\hline
	\end{tabular}
\caption{This table shows the expected number of events for a b-tagged study of
	$b \bar{b} h$ at 8 TeV for 20 $\fbinv$ of luminosity and a $\Lambda = 6
	~\tev$ cutoff. In addition to cuts (1)-(3) given in subsection
	\ref{subsec:jjh}, it was additionally required that the vector sum of
	the diphoton $p_T$ be greater than 150 GeV and that one of the two
	highest $p_T$ jets pass the medium charged secondary vertex b-tagging
	requirements detailed in \cite{Chatrchyan:2012jua}. Cuts (4)-(7) in
	subsection \ref{subsec:jjh}, which target VBF-produced Higgs bosons,
	were not applied to these events. The relative rate after collider cuts,
	$\mu_{b \bar{b} h}^{\text{collider}}$, can be compared to the tree-level
	parton cross-section $\mu_{b \bar{b} h}^{\text{parton}}$, to see the
	efficacy of the $p_T>150~\gev$ cut on the vector summed photons.}
\label{tab:hbb}
\end{table}

Comparing the results in Table \ref{tab:hbb} to those in Table \ref{tab:hjj}, we
find that adding a single $b$-tag and requiring the vector summed photon $p_T > 150 ~\gev$ increases the expected significance of detection
or exclusion of the chromomagnetic dipole operator. For a $c_{ghd} = 40$
coupling and a $\Lambda = 6\, \tev$ cutoff, and assuming a Poisson distribution with 0.65 SM events expected and a $\pm 0.5$ events systematic error, we find an $O(2\,
\sigma)$ exclusion could be obtained with the current 8 TeV data. As already
discussed, if an excess of events were found indicating $c_{ghd} = 40$ for
$\Lambda = 6 ~{\rm TeV}$, this would indicate new physics at scales around a
$\tev$, meaning the new physics responsible would have to avoid resonant $b
\bar{b}$ production that dijet resonance searches have excluded for $\Lambda \sim 1 ~{\rm TeV}$. In Table \ref{tab:hbb} we also give the parton-level relative cross-section rate $\mu_{b \bar{b} h}^{\text{parton}} \equiv \sigma/\sigma_{NP}~ |_{\rm{Parton~Level}}$ for SM vs new physics events. We remind the reader that $\mu_{b \bar{b} h}^{\text{parton}}$ does not include the affect of kinematic cuts, while $\mu_{b \bar{b} h}^{\text{collider}}$ does. The efficacy of the collider cuts on photon transverse momentum can be observed by comparing $\mu_{b \bar{b} h}^{\text{parton}}$ to $\mu_{b \bar{b} h}^{\text{collider}}$ in Table \ref{tab:hbb}.

While a $b$-tagged search for non-standard $b \bar{b} h$ rates and kinematics has not yet been conducted, there is a set of inclusive Higgs measurements with $b$-tags worth comment, namely searches for
MSSM (or 2HDM) Higgses. 
These searches are typically in the $pp(\bar p) \to
b\bar b \Phi(\bar b b)$~\cite{Behr:2013ji} or $b\bar b \Phi(\tau^+ \tau^-)$~\cite{CMS-PAS-HIG-12-050,Aad:2011rv} channels, where $\Phi$ can
stand for any one of the three neutral Higgses in a 2HDM.  These final states are
chosen to take advantage of the increased bottom quark and tau lepton Yukawa
couplings when $\tan{\beta}$ is large. However, the bulk of the sensitivity of
these searches comes when the second set of neutral Higgses (mass $m_A$) are
light. To understand this, recall that in the decoupling limit ($m_A \to
\infty)$ the bottom-quark coupling to the light SM Higgs asymptotes to the SM
value. The couplings that are actually enhanced are the couplings of the bottom
quarks (taus) to the heavy Higgses, therefore the enhancement in MSSM/2HDM
$pp(\bar p) \to b\bar b \Phi(\bar b b)$ rates falls off as $m_A$ grows and the
heavy states are less abundantly produced. We see from Table \ref{tab:hbb} that a chromomagnetic dipole coupling of $c_{ghd}=40$ only doubles the expected rate of $b \bar{b} h$ events, which corresponds to $\tan{\beta} \sim \sqrt 2$, a small $\tan{\beta}$ outside the sensitivity of MSSM/2HDM studies. Hence, recasting MSSM/2HDM $pp(\bar
p) \to b\bar b \Phi(\bar b b)$ limits in terms of $\mathcal O_{ghd}$, does not
lead to a strong bound. Another way to understand this lack of sensitivity is that, while these MSSM searches
look for $\Phi \to bb$, they don't impose a boost requirement on the $\Phi$.
With no boost requirement, the cross section is dominated by the low-$p_T$
region, where $\mathcal O_{ghd}$ has little effect.  

\subsection{Constraints on $b \bar{b} h$ at the 14 TeV LHC}

Of course, sensitivity to $\mathcal{O}_{ghd}$ via
$b \bar{b} h$ can be improved upon with a high luminosity run at the 14 TeV LHC. As we have the luxury of higher energy and luminosity, we can cut harder on the Higgs $p_T$ than at $8\,\tev$, obtaining a better signal to background ratio at the expense of overall rate. In particular, we studied the sensitivity of the 14 TeV LHC to the chromomagnetic bottom quark operator using the following cuts:
\begin{itemize}
	\item Exactly two photons and two or more jets, all satisfying the same same identification criteria (cuts (1)-(3)) as in subsection \ref{subsec:jjh}.
	\item The two highest $p_T$ jets must pass the tight charged secondary vertex b-tagging requirements (55\% tag, 0.1 \% light quark mis-tag) detailed in \cite{Chatrchyan:2012jua}. Whereas at 8 TeV, it was necessary to use the medium CSV b-tag to keep as many $b \bar{b} h$ events as possible, at 14 TeV we find that removing the $jjh$ background yields better significance. Note that for both the medium (light) CSV b-tags, we match the results of \cite{Chatrchyan:2012jua} by assuming that 20\% (10\%) of charm quarks will fake bottom quark jets. 
	
	\item The vector sum of diphoton $p_T$ must be greater than $200$ GeV.
		This cut tends to exclude non-Higgs produced $b \bar{b} \gamma
		\gamma$ backgrounds, and further discriminates between an
		un-boosted Standard Model sample of $b \bar{b} h$ events, and
		events with Higgs bosons boosted by the presence of a
		chromomagnetic dipole operator. This selection for boosted Higgs
		events can be seen by comparing the ratio of cross-sections $\mu^{\text{parton}}$,
		given in Table \ref{tab:14_TEV} to the ratio of selected events, $\mu^{\text{collider}}$. For example, looking at $c_{ghd}=20$, we see that the cross-section is around $40\%$
		larger than in the SM, but the number of expected events after the cut on vector diphoton $p_T$ is applied is 15 times
		larger than the SM expectation.
		\item As with the previous analysis, all cuts tailored to VBF
			production (cuts (4)-(7) from subsection
			\ref{subsec:jjh}) are dropped.
\end{itemize}

Repeating the same signal and background generation and analysis chain as in the
previous section and assuming $3\,\abinv$ of
integrated luminosity, we find the following event counts (Table \ref{tab:14_TEV}).
\begin{table}[h!]
	\renewcommand{\arraystretch}{1.4}
	\begin{tabular}{|c||c||c||c||c|c|c|c|c|c|c|}
		\hline
		& $ j j \gamma \gamma$ bkgd & $ b \bar{b} \gamma \gamma$ bkgd & $c_{ghd} =$ & -20 & -10 & -5 & 0 & 5 & 10 & 20 \\
		\hline
		14 TeV $b \bar{b} h$ events ($3\, \abinv$) & 0.1 & 7.5 & & 48 & 13 & 4 & 3 & 5 & 14 & 49  \\
		\hline
		14 TeV $\mu_{b \bar{b} h}^{\text{collider}}$ & -- & -- & & 16 & 4.3 & 1.3 & 1 & 1.7 & 4.7 & 16  \\
		\hline
		14 TeV $\mu_{b \bar{b} h}^{\text{parton}}$ & -- & -- & & 1.72 & 1.22 & 1.08 & 1 & 0.99 & 1.06 & 1.41  \\		
		\hline
	\end{tabular}
	\caption{The number of expected events are shown for indicated $c_{ghd}$ couplings and a $\Lambda = 6 ~\tev$ cutoff at the 14 TeV LHC after $3\,
		\abinv$, after applying cuts (1)-(3) detailed in subsection
		\ref{subsec:jjh}, in addition to requiring that the diphoton
		$p_T$ vector sum exceed 200 GeV and that both of the highest
		$p_T$ jets pass the tight charged secondary vertex b-tagging
		requirement (55\% tag, 0.1\% mis-tag) detailed in
		\cite{Chatrchyan:2012jua}. The relative rate $\mu_{b \bar{b} h}^{\text{parton}}$ of the new physics vs. SM $b \bar{b}h$ parton-level cross-sections can be compared to the collider-level event ratio $\mu_{b \bar{b} h}^{\text{collider}}$ to quantify the affect of the $p_T > 200~\gev$ cut on vector summed diphoton transverse momentum. Thus the selection for new physics events is a result of both cross-section rate and the non-standard kinematics of the chromomagnetic dipole.}
	\label{tab:14_TEV}
\end{table}

From Table \ref{tab:14_TEV} we see that an $O(2 \sigma)$
exclusion of $c_{ghd}=5$ could be achieved after 3
$\rm{ab}^{-1}$ luminosity\footnote{Here we simply calculate significance as the signal divided by the
square root of the expected SM background.}. This corresponds to a shift in the effective cutoff
from a TeV up to roughly 6 TeV -- implying sensitivity to half the LHC's energy
over its lifetime. Note that the ability to probe down to a $\Lambda \sim 6
~\tev$ cutoff hinges on the cut on the vector summed diphoton $p_T > 200 ~\gev$
-- this cut preferentially selects boosted events, which will be abundantly
produced if the $b$ quark has a chromomagnetic dipole. A coarse scan over possible
values found that a cut of $p_T > 200 ~\gev$ produced the highest ratio of new
physics ($c_{ghd} = 5$) to SM events without driving both to zero; the separation between signal and background increases with the $p_T$ cut, but the higher the cut value, the lower the overall rate. In summary, we find that the structure of the $b$ quark chromomagnetic operator can be thoroughly structinized at $\sqrt{s}=14 ~\tev$ and high luminosity.

The numbers in Table~\ref{tab:14_TEV} were derived assuming the only Higgs decay
mode viable for Higgs plus dijet searches is the $\gamma\gamma$ mode. This is a
safe, though somewhat conservative assumption. Expanding the search to more
decay modes will, in principle, improve the reach, though in practice many other
decay modes seem quite challenging to observe: $h \to b \bar b$ has the largest
signal rate but must compete with immense QCD backgrounds; $h \to WW^*$ -- once
combined with the accompanying $b\bar b$ -- forms the identical final state to
$t\bar t$ production; and $h \to \tau^+\tau^-$ events reconstruct the Higgs too
inefficiently once one takes into account the difficulty in identifying taus.
It is possible that multivariate techniques could find enough kinematic
differences between these `alternate' signals, i.e. $pp \to h(b\bar b) +
b\,\bar b$, and the SM background, but dedicated studies beyond the scope of this
paper are required.

\section{Conclusions}
\label{sec:conc}
Since it is the heaviest quark that hadronizes, and is the easiest to identify
at the LHC, the bottom quark opens a unique door onto the
vista of new Higgs dynamics. Sensitivity to the chromomagnetic dipole, which has
a distinct momentum structure resulting in a non-standard fraction of boosted
Higgs bosons,
is paramount in the ongoing quest for new physics. In this work we have pinpointed the effects of this single dimension-6 effective operator. A generic UV theory will likely produce several other dimension-6
operators after heavy degrees of freedom are integrated out. Considering the
effects of other operators would change the exact numbers of new physics events
quoted here, but the efficacy of a search centered on the unique kinematic structure 
induced by the chromomagnetic dipole would remain unchanged. To wit, searches for these other operators have a scope restricted to the overall rate of $b \bar b h$ production. A focused search for the chromomagnetic dipole boosting Higgs $p_T$ allows us to probe to higher energy scales: in the case of this study, we find the chromomagnetic dipole could be probed up to effective energy scales of roughly  6 TeV over the course of a high luminosity LHC run. 

In this paper, we first cataloged what bounds can be put on the $b$ quark
chromomagnetic dipole from processes $h \rightarrow b \bar b$, $pp \rightarrow
h$, $pp \rightarrow b \bar b$, and $p p \rightarrow h q \bar q$. There are no
bounds on the chromomagnetic operator coming from inclusive decays of the Higgs
to $b \bar{b} g$ (less than a thousandth the total width), or loop production of
the Higgs; the amplitude for this latter process is suppressed by a factor of
the bottom Yukawa stemming from a chirality flip. We have found that $pp\to
b\overline{b}$ constrains the $b$ chromomagnetic dipole Wilson coefficient
$c_{ghd} < 40$ for a 6 TeV cutoff. On the other hand, presently available
collider studies of Higgs production in association with two jets -- which do not employ $b$-tagging -- do not constrain the $b$ quark chromomagnetic dipole. 

Casting an eye towards future discovery prospects for $b \bar b h$, we have found that the sensitivity of Higgs plus dijet searches to the $b$ quark chromo-dipole operator at 8 TeV with $19.6\; \text{fb}^{-1}$ and particularly at 14 TeV with $3\, \abinv$ will be greatly increased with the addition of $b$-tagging in addition to cuts on transverse momentum in $j j \gamma \gamma$ final states. Both 8 and 14 TeV LHC studies of the $jj h$ channel should include $b$-tags to possibly catch the first glimmer of new physics in dijet plus diphoton ($m_{\gamma \gamma} \sim 125 \rm ~GeV$) events. 

The effect of the bottom quark chromomagnetic dipole is one 
example where an upgrade of the LHC to high luminosity will be very beneficial. 
We have demonstrated that in order to exclude an order one coupling
chromomagnetic operator we will need $3\,\abinv$ of luminosity, which clearly shows that more
statistics will play an essential role in understanding Higgs precision physics. Indeed, it should not be surprising that in order to exclude $\sim 6 ~\tev$ new physics coupled to the Higgs and bottom quarks at a 14 TeV machine, a high luminosity is required. The HL-LHC is a
natural arena to search for new operators that affect Higgs physics
not by changing its properties at its pole mass but by changing the kinematic distributions
at high momentum transfer. 
\\

\textbf{Acknowledgements} 
We thank Kevin Lannon, Bryan Ostdiek, and Anna Woodard for useful conversations. 
This work was supported in part by the Notre Dame Center for Research 
Computing through computing resources. The work of AD was partially supported by the National Science Foundation under Grant No. PHY-1215979, and the work of AM was partially supported by the National Science Foundation under Grant No. PHY14-17118.

\bibliographystyle{JHEP}

\bibliography{bbh}

\end{document}